\documentclass[12pt]{article}
\usepackage{psfig}
\def\TL{\hfil$\displaystyle{##}$}
\def\TR{$\displaystyle{{}##}$\hfil}

\def\TT{\hbox{##}}
\def\seqalign#1#2{\vcenter{\openup1\jot
  \halign{\strut #1\cr #2 \cr}}}

\def\comment#1{}
\def\fixit#1{}

\def\tf#1#2{{\textstyle{#1 \over #2}}}

\def\mop#1{\mathop{\rm #1}\nolimits}

\def\coth{\mop{coth}}

\def\sgn{\mop{sgn}}

\def\overleftrightarrow#1{\vbox{\ialign{##\crcr
     $\leftrightarrow$\crcr\noalign{\kern-0pt\nointerlineskip}
     $\hfil\displaystyle{#1}\hfil$\crcr}}}

\def\lsim{\mathrel{\mathstrut\smash{\ooalign{\raise2.5pt\hbox{$<$}\cr\lower2.5pt\hbox{$\sim$}}}}}
\def\gsim{\mathrel{\mathstrut\smash{\ooalign{\raise2.5pt\hbox{$>$}\cr\lower2.5pt\hbox{$\sim$}}}}}

\def\sqr#1#2{{\vcenter{\vbox{\hrule height.#2pt
         \hbox{\vrule width.#2pt height#1pt \kern#1pt
            \vrule width.#2pt}
         \hrule height.#2pt}}}}
\def\square{\mathop{\mathchoice\sqr56\sqr56\sqr{3.75}4\sqr34\,}\nolimits}

\def\href#1#2{#2}  

\def\lbldef#1#2{\expandafter\gdef\csname #1\endcsname {#2}}
\def\eqn#1#2{\lbldef{#1}{(\ref{#1})}%
\begin{equation} #2 \label{#1} \end{equation}}
\def\eqalign#1{\vcenter{\openup1\jot
    \halign{\strut\span\TL & \span\TR\cr #1 \cr
   }}}

\def\arcsinh{\mop{arcsinh}}

\def\d{{\nabla}}
\def\comment#1{  \begin{raggedright}{\tt [#1]}\end{raggedright}}
\begin{document}
\baselineskip=14.5pt
\pagestyle{plain}
\setcounter{page}{1}
\renewcommand{\thefootnote}{\fnsymbol{footnote}}
\newcommand{\da}{\dot{a}}
\newcommand{\db}{\dot{b}}
\newcommand{\dn}{\dot{n}}
\newcommand{\dda}{\ddot{a}}
\newcommand{\ddb}{\ddot{b}}
\newcommand{\ddn}{\ddot{n}}
\newcommand{\pa}{a^{\prime}}
\newcommand{\pb}{b^{\prime}}
\newcommand{\pn}{n^{\prime}}
\newcommand{\ppa}{a^{\prime \prime}}
\newcommand{\ppb}{b^{\prime \prime}}
\newcommand{\ppn}{n^{\prime \prime}}
\newcommand{\fda}{\frac{\da}{a}}
\newcommand{\fdb}{\frac{\db}{b}}
\newcommand{\fdn}{\frac{\dn}{n}}
\newcommand{\fdda}{\frac{\dda}{a}}
\newcommand{\fddb}{\frac{\ddb}{b}}
\newcommand{\fddn}{\frac{\ddn}{n}}
\newcommand{\fpa}{\frac{\pa}{a}}
\newcommand{\fpb}{\frac{\pb}{b}}
\newcommand{\fpn}{\frac{\pn}{n}}
\newcommand{\fppa}{\frac{\ppa}{a}}
\newcommand{\fppb}{\frac{\ppb}{b}}
\newcommand{\fppn}{\frac{\ppn}{n}}
\newcommand{\A}{A}
\newcommand{\B}{B}
\newcommand{\mmu}{\mu}
\newcommand{\mnu}{\nu}
\newcommand{\ii}{i}
\newcommand{\jj}{j}
\newcommand{\jl}{[}
\newcommand{\jr}{]}
\newcommand{\ml}{\sharp}
\newcommand{\mr}{\sharp}



\begin{titlepage}

\begin{flushright}
MIT-CTP-3099 \\
hep-th/0011156
\end{flushright}
\vfil

\begin{center}
{\huge Locally Localized Gravity }\\[8pt]
\end{center}

\vfil
\begin{center}
{\large Andreas Karch and
Lisa Randall$^*$}
\end{center}

$$\seqalign{\span\TL & \span\TT}{
 & Center for Theoretical Physics,
  \cr\noalign{\vskip-1.5\jot}
   & Massachusetts Institute of Technology, Cambridge, MA  02139-4307, USA
  \cr\noalign{\vskip0.5\jot}
}$$
\vfil

\begin{center}
{\large Abstract}
\end{center}

\noindent

We study the fluctuation spectrum of linearized gravity around 
non-fine-tuned  branes. 
We focus on the case of an AdS$_4$ brane in AdS$_5$. In this
case, for small cosmological constant, the warp factor near the
brane is essentially that of a Minkowski brane. However, far from
the brane, the metric differs substantially. The space includes
the AdS$_5$ boundary, so it has infinite volume. 
Nonetheless,  for sufficiently small
AdS$_4$ cosmological constant, there is
 a bound state graviton in the theory,
and four-dimensional gravity is reproduced. However,
it is a massive bound state that plays the role of the
four-dimensional graviton.

\vfil
\begin{flushleft}
November 2000
\end{flushleft}

{\vskip 5pt \footnoterule\noindent
{\footnotesize $\,^*$\ {\tt 
karch@mit.edu, randall@.mit.edu}}}
\end{titlepage}
\newpage
\baselineskip=15.5pt

\section{Introduction}
The realization 
that even spin 2 excitations can be localized by a gravitating brane 
\cite{RS2} suggests an alternative to compactification,
since one has an infinite extra dimension but nonetheless
sees four-dimensional gravity at low scales. However, it
seemed a critical aspect of this proposal was the finite
``volume'' of the extra dimensional space 
since that is
what permitted a normalizable zero-mode and a finite
value of the four-dimensional Planck scale. However,
it would be perplexing if this finite volume is truly essential,
since we should expect physics to be local,
so it would be strange if there were strong
dependence on the warp factor far from
 the brane on which gravity potentially localizes.

For this reason, it is very interesting to study an
AdS$_4$  brane in AdS$_5$, in which the tuning of
cosmological terms required for a flat brane is modified.
The metric is known in this case, and it is a warped
geometry in which the warp factor blows up at large distance
from the brane, rather than asymptoting to zero as it does
in the flat brane case. Therefore, the geometry includes
the boundary of AdS$_5$ and the volume of the extra-dimensional
space is infinite. Clearly, the zero mode, corresponding
to a perturbation whose wavefunction in the 5th dimension
is proportional to the warp factor, is nonnormalizable.

Nonetheless, since such a theory can be obtained from
the flat brane scenario through an arbitrarily small
reduction of the cosmological term on the brane, it
is difficult to believe that there is no bound state
graviton to generate four-dimensional physics. Indeed,
we find that four-dimensional gravity is reproduced,
due to a very light, but massive bound state Kaluza-Klein mode
of the graviton. Of course the gravity we reproduce is
AdS gravity, so the cosmological constant would have to be
sufficiently small so that physics looks Minkowskian within the horizon size.

This is an important result, as it demonstrates
that localization is  a local phenomenon, as indeed it must be.
A warped geometry in a local region can be sufficient
to generate four-dimensional gravity independently
of the behavior of the higher dimensional geometry far away.
Although we explicitly analyze the AdS$_4$ brane
in AdS$_5$, we expect this result is far more general.

For those familiar with the c-theorem in five
dimensions, it might be surprising that we can have
a warp factor that localizes gravity near the brane,
but then rises again. However, we will show
this is consistent in AdS space.

This theory probably has interesting implications
from the AdS/CFT vantage point, since
it includes both a Planck brane and the AdS$_5$ boundary,
which we  briefly comment on here.

Furthermore, this theory
gives rise  to four-dimensional  gravitational physics on a brane embedded
into an AdS$_5$ bulk that asymptotes to the boundary of AdS and not
the interior as its Minkowski or dS cousin, thereby avoiding
the no-go theorems about smooth RS branes and supergravity and possibly
faciliating a realization of localized gravity in that context.

In Section 2, we  review the background geometry for a single
brane with a non-finetuned tension, that is with an effective
cosmological constant. In Section 3 we will discuss the fluctuation
spectrum around such a background. While in the dS
case one still has a massless bound graviton, we 
will show that in the AdS case, the massless bound state is lost. 
However one does find an almost massless graviton bound on
the brane, which we refer to as the almost zero mode.
In Section 4, we argue that four-dimensional
gravity is correctly reproduced.
We conclude in the final section.

\section{The Background}
\subsection{Action and Background Solution}

As in the original  setup \cite{RS2},
the action we consider is just that of 5d gravity with a negative cosmological
constant $\Lambda^{5d} = -3/L^2$ coupled to a brane of tension $\lambda$:

\eqn{SGRAV}{S = \int d^5x \, \sqrt{g} \left[ -{1 \over 4} R -
   \Lambda^{5d} \right] \ - \lambda \, \int d^4x \,
dr \, \sqrt{| \det g_{ij}|} \, \delta(r ),}
where $g_{ij}$ is the metric induced on the brane by the ambient
metric $g_{\mu\nu}$.

We use the ansatz for the solution to be a warped product
with warp factor $A(r)$,
\eqn{deSitterAnsatz}{
   ds^2 = e^{2 A(r)} \bar{g}_{ij} dx^i dx^j - dr^2 \ ,
  }
allowing for the 4d metric to be Minkowski, de Sitter or anti-de
Sitter with the 4d cosmological constant $\Lambda$ being
zero, positive or negative respectively following the conventions
of \cite{dWFGK}.

The solutions to Einstein's equations\footnote{Actually
these solutions have been known and studied before, \cite{cvetic1,cvetic2}.
It is in the references we give that they have
been reconsidered in the context of localized gravity.}
are \cite{dWFGK,Kaloper,KimSquared,
Nihei}:
\eqn{deSitterWall}{\eqalign{
    dS_4 :
   & A = \log (\sqrt{\Lambda} L \sinh {c -| r| \over L}) \ ,
\quad \lambda = {3 \over L} \coth {c \over L}  \cr
    M_4 : & A = {c - |r| \over L} \ ,
\quad \lambda = {3 \over L}  \cr
    AdS_4 :
   &  A = \log (\sqrt{-\Lambda} L \cosh {c - |r| \over L}) \ ,
\quad \lambda = { 3 \over L} \tanh {c \over L} \ .
  }}

The geometry of the curved solutions is characterized by a 
parameter $c$, which is determined in terms of the brane tension.
In the dS case $c$ is the distance between the  brane and the horizon,
whereas
in the AdS case $c$ is the distance to the turn around point
in the warp factor.
As is well known, in order to have a Minkowski solution,
one has to fine-tune $\lambda$ relative to $L$; 
in our conventions $\lambda = {3 \over L}$. Since $\coth(x) > 1$
and $\tanh(x) < 1$ for any $x$, we see that,
$\lambda > {3 \over L}$ implies that the effective
4d cosmological constant is positive and $\lambda < {3 \over L}$
implies a negative 4d cosmological constant. The 4d cosmological constant
in 4d Planck units is given by
\eqn{BigTune}{
   4\pi \int dr \, e^{2 \left( A(r) - {1 \over 2} \ln|\Lambda|
    \right)} 
  }
and is hence determined by $c$ alone, that is by the mismatch of
brane tension and bulk cosmological constant.
Since
only the combination $\Lambda e^{2A}$ appears
in the equations of motion, not both, $\Lambda$ and $A$, will
be determined independently. One can use
this freedom to set $A(0)=0$, as in \cite{RS2}. This way the
cosmological constant becomes
\eqn{cc4d}{ \Lambda_{dS} = \frac{1}{L^2 \, \sinh^2{\frac{c}{L}}},  \; \;\; \;
 \Lambda_{AdS} = \frac{-1}{L^2 \, \cosh^2{\frac{c}{L}}}. }
Using the value of $c$ determined by the jump equations one finds
that the 4d cosmological constant
 is indeed only given by the detuning 
\eqn{mismatch}{M=\frac{\lambda L}{3}}
of the brane tension 
without any exponential surpression:
\eqn{cc4d2}{ \Lambda_{dS} = \frac{1}{L^2} (M^2-1),  \; \;\; \;
\Lambda_{AdS} = \frac{1}{L^2} (1-M^2). }

 \begin{figure}
   \centerline{\psfig{figure=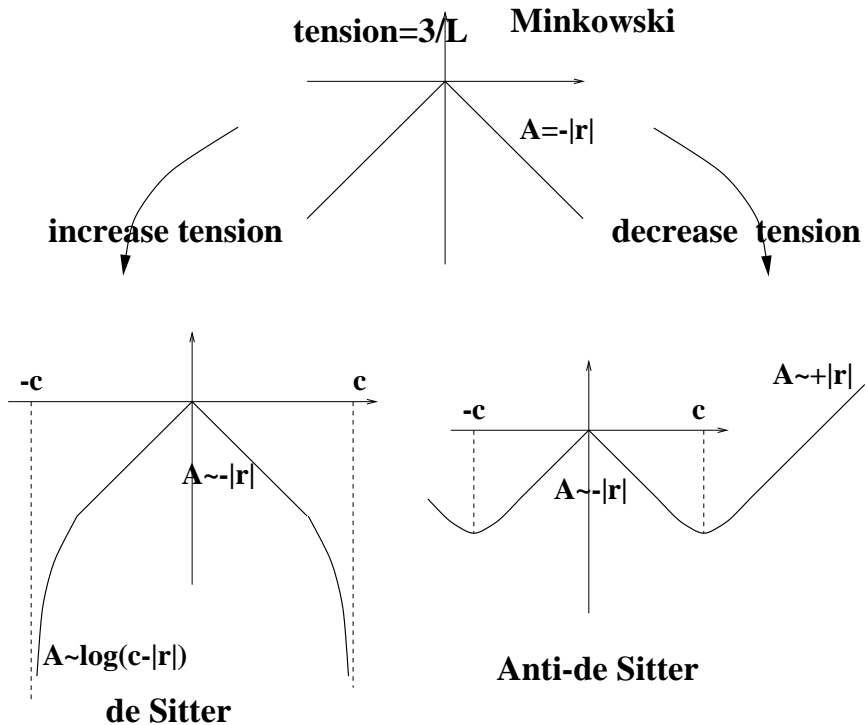,width=4.5in}}
    \caption{The behavior of the warp-factor for $\Lambda=-1$, 0 and 1.}
\label{warp}
  \end{figure}
The general behavior of the three solutions is sketched in figure~\ref{warp}.
In the fine-tuned Minkowski solution, the warp factor is just
$A=-|r|$ and the graviton is marginally bound on the brane; that
is, it is a bound state at threshold. The onset of the continuum
is not separated by a mass gap. Increasing
the tension the backreaction on the warp factor is even stronger
and even though $A(r)$ starts out linear in $r$, as in
the Minkowski case, it goes to $-\infty$ at a finite value of $r$, at $r=c$.
$A(r)$ diverges as $\log(c-|r|)$,
but all components of the curvature tensor stay 
finite.
Since the bulk solution is nothing but
AdS$_5$, the coordinate singularity we see in this particular
coordinate system is just a horizon.
Gravity  is still trapped in the usual way.
Since we increased the tension of the brane, one should expect
that the marginal bound state becomes a true bound
state.
Formally one can
even take the limit where the bulk cosmological constant vanishes altogether,
$L \rightarrow \infty$, in which case one still traps gravity.
The curvature vanishes identically in the bulk. The coordinate singularity
reduces to the Rindler horizon of an accelerated observer in flat Minkowski
space; see \cite{Kaloper}.

If we however decrease the tension of the brane as compared to the
fine-tuned value, the backreaction on the warp factor becomes ``weaker''.
Even though it still starts out as $A\sim-|r|$ close to the brane
it reaches a minimum at a $r=c$ and then turns around and
runs toward $+\infty$ as $A\sim +|r|$, approaching
the boundary of AdS$_5$ and not the interior. So even though
close to the brane it still looks like we trap gravity in the standard
way, much of the amplitude for the zero mode state is
concentrated near the boundary of the bulk space. However,
since the cause of losing the bound graviton arises
only far from the brane, at what should correspond
to low energies, one should expect localization.
In the next section, we will show that four-dimensional
gravity is reproduced, first by a propagator analysis,
and subsequently,  by showing the interesting mode
analysis that gives the same result.  The
authors of \cite{dWFGK} gave the warp factor but
 did not study this physics, presumably
due to the absence of phenomenologists.

\subsection{The Modified C-theorem}

In the language of AdS flows
the positive energy condition has been rephrased as the so called
``c-theorem'' \cite{fgpw1,girardello} 
which asserts that $A'' \leq 0$. This is one of the reasons that
it was believed that localization of gravity on a positive
tension brane goes together with a warpfactor that tends
asymptotically to the AdS horizon, since it can not turn around.
In the solution we are considering $A''$ is obviously positive
even though our brane has positive tension, so it does not violate
positivity of stress energy.
Let us see how this is consistent with the c-theorem.

The derivation
of the c-theorem required Lorentz invariance and hence 
is only valid for Minkowski solutions.
Even the AdS$_4$ brane violates $A'' \leq 0$, since
\eqn{adoubleprime}{
A''(r) = {1 \over L^2} {1 \over \cosh^2 {c \pm |r| \over L}}
}
is positive everywhere in the bulk. 
Using the fact that the energy
momentum tensor supporting our geometry
has to be of the form
\eqn{energymomentum}{
T^{\alpha}_{\beta} = \mbox{diag} \{ \rho, -p_1, -p_2, -p_3, -p_4 \} 
}
what is called the "weaker energy condition" in \cite{fgpw1}
\eqn{weaker}{
T_{\alpha \beta} \xi^{\alpha} \xi^{\beta} \geq 0 }
where $\xi^{\alpha}$ is an arbitrary null vector translates into the
condition
\eqn{conditiononmatter}{
\rho + p_i \geq 0. }
It is easy to see that energy momentum tensors supporting
solutions of the general form of a warped
product based on 4d Minkowski, de Sitter or anti-de Sitter space
will automatically satisfy the conditions for $i=1,2,3$. Since
\eqn{Einstein55m00}{
-3 A'' - 3 \Lambda e^{-2A} = R^0_0 - R^5_5 = G^0_0 -G^5_5 = (T^0_0
- T^5_5) \geq 0}
the constraint on $A''$ from the weaker energy condition reads
\eqn{genCtheorem}{
A'' \leq -  \Lambda e^{-2A}.}
For positive $\Lambda$ this implies an even stronger version
$A'' <0$ of
the standard c-theorem, $A'' \leq 0$. However for negative $\Lambda$
in general positive $A''$ is allowed. The solutions written down above
actually saturate the bound.

\section{Spectrum of Linearized Fluctuations}

\subsection{Gravity on the AdS brane}

Despite the absence of a massless graviton, we expect
 physics to be continuous and to react smoothly
to small changes in the input parameters, like the brane tension. If we were
to lose  localized gravity by going
from zero to tiny negative cosmological constant, this would mean that by
the very fact that four-dimensional gravity exists, we could rule out
 an AdS$_4$ brane world.
Similarly, one can argue that low energy
physics on the brane should only probe the warp factor close to the brane,
which is basically indistinguishable for the Minkowski and AdS cases,
and hence should  be insensitive to the eventual turn around
 in the warp factor. From a holographic perspective,
if we probe sufficiently short distances on the brane,
physics should not depend on the warp factor far away.
This fact becomes even more striking when we heat up the brane universe
to finite temperature. In this case a black hole will form in the bulk,
and space time gets cut off at the black hole horizon, manifestly
localizing gravity and hiding all the differences
between the zero and negative cosmological constant scenario
behind an event horizon.

Indeed, we argue that four-dimensional gravity is localized
to the brane. We first argue that the propagator
between two points that are sufficiently close relative
to the scale set by the AdS$_4$ curvature will reduce
to the Minkowski brane propagator.

In Minkowski space,  
one finds \cite{katz} that 
in the conformal ``z-coordinates'' in which the 
metric reads $ds^2=e^{2A(z)}(g_{ij}dx^i dx^j-dz^2)$ 
the propagator can be written as
\eqn{propfourier}{
\Delta_{d+1} ( x,z;x',z') = \int \frac{d^4p}{(2 \pi)^4} e^{i p (x-x')}
\Delta_p(z,z')}
where
\eqn{equationinmom}{
e^{-2A} ( \partial_z^2 + 3 A' \partial_z - p^2) \Delta_p(z,z') =
e^{-5A} \delta(z-z'). }
In the AdS case, the integral in the Fourier transform will be
replaced by a sum over discrete modes, whose spacing is of
order $|\Lambda|$. So unless we want to study cosmological
distance scales, the sum can still be approximated by an integral and
then \equationinmom\ tells us that the propagator for processes
on the brane only
depends on the warp factor in the vincinity of the brane.

Performing the coordinate change to the
``z-coordinate'', as we will work out later, the warp factor reads
\eqn{tooearly}{
e^A=\frac{L \sqrt{|\Lambda|} }{\sin {\left ( (|z| + z_0) \sqrt{|\Lambda|}
\right ) }}.}
For small $|z|$, the $\sin$ can be expanded and we obtain
to leading order in $|z|$
\eqn{approxwarp}{
e^A = \frac{1}{1+ \cos{\left ( z_0 \sqrt{|\Lambda|}
\right ) }\; \; |z|/L},}
which is identical to the flat space form $e^A=\frac{1}{|z|/L +1}$
up to a rescaling in $L$, so that the form of the differential
equation stays unchanged. The higher order corrections are
accompanied with powers of $|\Lambda| \sim e^{-2c}$ and hence
are very small for near critical tension.

To complete this analysis, and demonstrate
the correct polarization structure of the propagator,
one would have to incorporate the effect of brane
bending. This will be discussed in the mode analysis
below. However, one can use the propagator analysis,
as performed in Ref. \cite{giddingskatz}, to see that
the correct gravitational potential is reproduced.

In  \cite{giddingskatz}, one explicitly solves for the trace
and the longitudinal components of the metric perturbation in terms of
the sources (rather than
changing coordinates so that
the brane is bent). 
These solutions then set up an ``effective source term''
for the TT-components, which includes contributions
from the trace and longitudinal components of the metric. In terms of the
trace reversed metric perturbation 
$\bar{h}_{\mu\nu}= h_{\mu\nu}- \frac{1}{2} h \eta_{\mu\nu}$ this effective
source with
stress energy only  on the brane 
leads in flat space leads to a linearized response
\eqn{branesol}{
\bar{h}_{\mu\nu} (x) = -{1\over 16} \, 
\int \, d^4 x^{\prime} \Biggl\{
\, \Delta_{5} (x,0;x^{\prime},0) T_{\mu\nu}(x') - \\
 \eta_{\mu\nu} \biggl[ \Delta_{5}
(x,0;x',0) - {2 \over L} \Delta_{4}(x,x')\biggr]  \,
{T^{\lambda}_{\lambda}(x') \over 4 }\Biggr\}
}
that contains
an extra piece  corresponding to the contribution of the
extra scalar, that cancels the
extra scalar contribution from the propagator
of a massive mode. A similar analysis works in our
AdS case  with the scalar picking up a mass $4 \Lambda$,
so that o ordinary gravity plus small corrections is reproduced.

It is important to see explicitly how the above propagator,
which was treated approximately in the above,  is
reproduced in a mode analysis. In the next
section, we solve for  the modes of linearized
fluctuations around the brane, so that one
can explicitly construct the propagator.

\subsection{Analog Quantum Mechanics for Transverse Traceless Modes}

In this section, we study the spectrum of linearized
gravity fluctuations around the dS$_4$ and AdS$_4$ solutions.
We first consider the transverse-traceless  (TT) modes, defined by
 \eqn{TT}{ D^j
h_{ij} = g^{ij} \, h_{ij} = 0 \,,  }
where the perturbed metric is
\eqn{PerturbedMetric} 
{ds^2 = e^{2 A(r)} (g_{ij} + h_{ij}) \, dx^i dx^j - dr^2
\,,}
and we have chosen axial gauge where
\eqn{AxialGauge}{ h_{\mu 5} = 0 \,,} with $\mu = 0,1,2,3,5$.
We will consider the non-TT modes in the next section,
and in more detail in \cite{withami}.
 As suggested
in \cite{RS2}, the TT modes are best analyzed by transforming Einstein's
equations for the transverse traceless modes of the
linearized fluctuations into the form of an
analog quantum mechanics, where they satisfy a standard
Schr\"odinger equation with a ``volcano potential'',
which can be derived for arbitrary warp factor, see e.g. \cite{
dWFGK,CsakiErlichHollowood}.

Their linearized equation of motion reads, see e.g. \cite{ortin,
garriga}
\eqn{TTEqn}{ \left( \partial_r^2 + 4 A' \partial_r - e^{-2A}
(\square_{4d} + 2 \Lambda) \right) h_{ij} = 0 \,.  }
We are looking for a mode corresponding to a 4d graviton,
or more generally, a spin 2 excitation of mass $m$; that is
\eqn{fourdgraviton}{
(\square_{4d} + 2 \Lambda) h_{ij} = m^2 h_{ij} }
where $\square_{4d}$ is the 4-dimensional covariant d'Alembertian
and $m^2$ the mass of the excitation.
In the curved case, the constant shift by $2 \Lambda$ is
precisely the one needed in order to recover the
equation of motion for a massless graviton (defined via the
reduced number of polarization states) 
for $m^2=0$.
Following the by now standard procedure, going to the conformal
``z-coordinate'' in terms of which the metric reads
\eqn{ConformallyFlat}{ds^2 = 
e^{2A(z)} \left(( g_{ij} + h_{ij}) \, dx^i dx^j - dz^2 \right) \,.}
and rescaling $H_{ij}(z) = e^{3A(z)/2} \, h_{ij},$ we obtain the
Schr\"odinger equation of the analog quantum mechanics
\eqn{SUSYQM}{
\left( - \partial_z^2 +  {9 \over 4} A'(z)^2 + 
{3 \over 2} A''(z) \right) H_{ij}(z) = m^2 H_{ij}(z) \,. }
with the volcano potential
\eqn{Volcano}{V(z) = \tf{9}{4} A'(z)^2 + \tf{3}{2} A''(z).}

\subsection{The Volcano Potentials for dS and AdS}

The change of coordinates to the conformally flat metric is given by
 \eqn{confflattrafo}{\eqalign{
   dS_4 : & \quad z(r) = \frac{1}{\sqrt{\Lambda}} \sgn(r) \;
\left \{ \arcsinh \left( 
\frac{1}{\sinh \left( \frac{c -|r|}{L} 
\right ) } \right) - z_0 \sqrt{\Lambda} \right \}
      \cr
   AdS_4 : & \quad z(r) = \frac{1}{\sqrt{-\Lambda}} \sgn(r) \;
\left \{ \arcsin \left( 
\frac{1}{\cosh \left( \frac{c -|r|}{L} 
\right ) } \right) - z_0 \sqrt{-\Lambda} \right \}
  }}
where the parameter $z_0$ is defined as
\eqn{zzero}{\eqalign{
   dS_4 :& \quad z_0 = \frac{1}{\sqrt{\Lambda}} \;  \arcsinh \left
  ( \frac{1}{\sinh \left( \frac{c}{L} \right ) } \right ) \cr
   AdS_4 :& \quad z_0 = \frac{1}{\sqrt{-\Lambda}} \;  \arcsin \left
  ( \frac{1}{\cosh \left( \frac{c}{L} \right ) } \right ). 
  }}
Note that while in the dS case the $z$ coordinate actually runs from 
$-\infty$ to $+\infty$ in the AdS case it only runs from $-(\frac{\pi}
{\sqrt{|\Lambda|}}-z_0)$ to
$(\frac{\pi}{\sqrt{|\Lambda|}} - z_0)$. 

The resulting volcano potentials are
\eqn{volcano}{\eqalign{
   dS_4 :& \quad V(z) = \frac{9 \Lambda}{4} + \frac{15}{4} \frac{
\Lambda}{
\sinh^2 \left ( \sqrt{\Lambda} (|z| + z_0) \right )} - 3 \;
\coth(\sqrt{\Lambda} z_0) \;
\sqrt{\Lambda} \; \delta(z) \cr
   AdS_4 :& \quad V(z) = \frac{-9 (-\Lambda)}{4} + \frac{15}{4} \frac{
(-\Lambda)}{
\sin^2 \left ( \sqrt{-\Lambda} (|z| + z_0) \right )} - 3 \;
\cot(\sqrt{-\Lambda} z_0) \;
\sqrt{-\Lambda} \; \delta(z)  . 
  }}

\subsection{The de Sitter Brane Spectrum }

 \begin{figure}
   \centerline{\psfig{figure=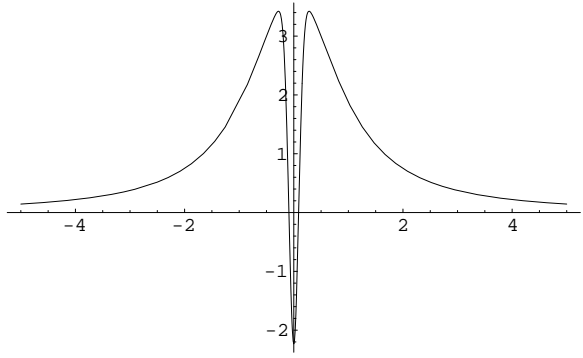,width=3.0in}}
    \caption{The Minkowski volcano potential.}
\label{min}
  \end{figure}
 \begin{figure}
   \centerline{\psfig{figure=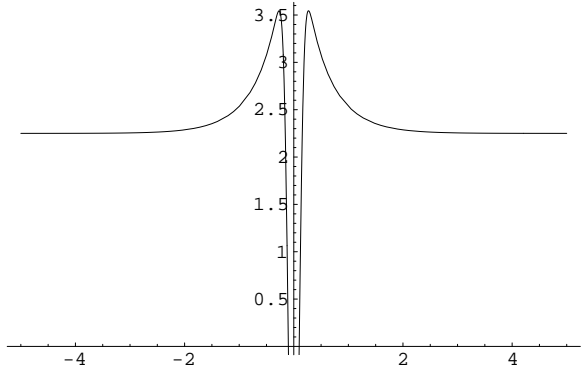,width=3.0in}}
    \caption{The deSitter volcano potential.}
\label{ds}
  \end{figure}
The volcano potential for the deSitter case is plotted in
Figure~\ref{ds}. The general features of this potential 
are very similar to those of the flat space volcano of
Figure~\ref{min}. Both potentials have a normalizable
zero mode trapped in the delta potential, whose wave function
is given by the warp factor $e^{\frac{3 A(z)}{2}}$ and decays like $1/z^3$.
In the Minkowski case this massless graviton is
a marginal bound state and the continuum of KK modes starts
at $m=0$. In the dS case, since at infinity the
potential approaches a constant value $\frac{9}{4} \Lambda$,
the onset of the continuum is only at $m^2=\frac{9}{4} \Lambda$
and the graviton is separated by a finite mass gap from the 
continuum \cite{garriga}.

\subsection{The Embedding of AdS$_4$}
 \begin{figure}
   \centerline{\psfig{figure=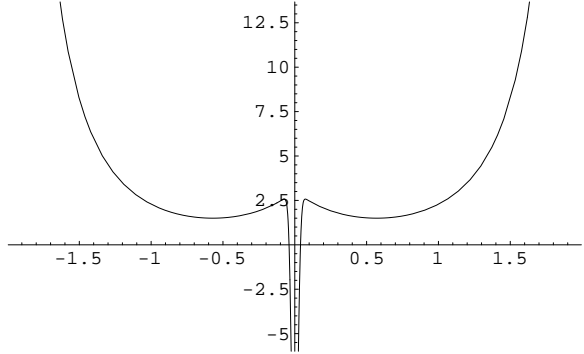,width=3.0in}}
    \caption{The Anti-deSitter volcano potential.}
\label{ads}
  \end{figure}
The Anti-deSitter volcano potential, as shown in Figure~\ref{ads},
looks quite different. As remarked above, the z-coordinate only
stretches over the finite range from $-(\frac{\pi}{\sqrt{|\Lambda|}} 
- z_0)$ to
$(\frac{\pi}{\sqrt{|\Lambda|}}-z_0)$. 
The potential diverges at the boundaries, so that we effectively
create a box. As we tune the effective 4D cosmological
constant from positive to negative values,
the flanks of the volcano move down from $\frac{9}{4} \Lambda$ to
zero and then
the potential ``curls'' in and we are left with a discrete
spectrum of modes without a zero mode. The would be zero mode
wavefunction diverges together with the warp factor $e^{A(z)}$ at
the boundaries of the box.

Before we solve for the discrete spectrum of the quantum mechanics explicitly
let us discuss why we see this behavior. The crucial
point is that while the region of space bounded by the
worldvolume of a single positive tension dS or
Minkowski brane does not contain any part of the boundary of AdS$_5$,
for the AdS brane this is not the case. 
 \begin{figure}
   \centerline{\psfig{figure=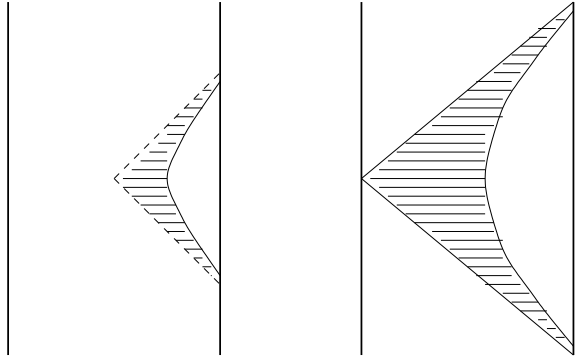,width=3.0in}}
    \caption{Schematics of the Penrose diagrams of the dS and Minkowski.
The spacetime one is instructed to keep is shaded. Since the
branes are accelerated, they have their own horizon, only for
the Minkowski brane does this coincide with the Poincare patch
horizon. The
spacetime one wants to keep is between the brane and the horizon.
The dS brane really is a full hyperboloid, the part drawn corresponding
to the static slice of dS. In
the case of the AdS brane the brane falls through the horizon
of the Poincare patch. It is more useful to study the embedding
in terms of an constant time slice through global AdS.}
\label{penrose}
  \end{figure}
\begin{figure}
   \centerline{\psfig{figure=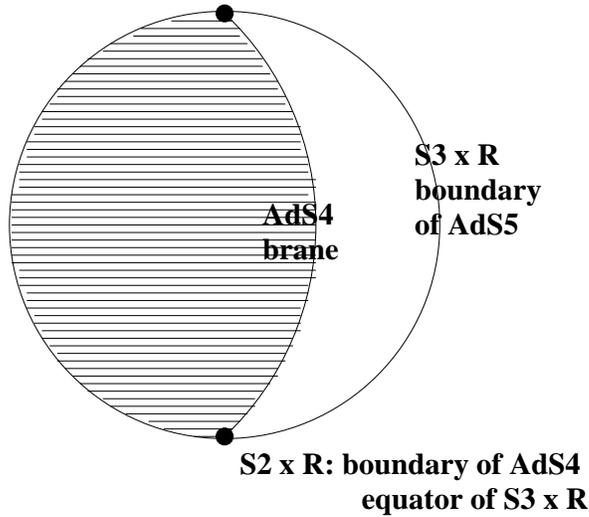,width=3.0in}}
\caption{Constant time slice through AdS$_5$, with the AdS$_4$ brane included
and again the region of spacetime we are instructed to keep shaded. Global
time on the brane is global time in AdS$_5$, the picture does not evolve in 
time.}
\label{slice}
  \end{figure}

Figures~\ref{penrose},~\ref{slice} summarize
the difference between the flat and deSitter
universes on  the one hand and the AdS universe on the other hand. 
For the former
the RS cutting
and pasting procedure leads to a space that ends in a horizon on both sides.
In the AdS$_4$ case
even though n the space-time
we keep it looks close to the brane like we are moving towards the horizon
(the warp factor decreases) eventually we have to go all the way out to the
boundary. While for the flat brane, gravitons sent out from the brane never
come back but fall through the horizon, for the AdS$_4$ brane some eventually
bounce back from the boundary and make it back to the brane. Effectively
the 5d spacetime is a box and the gravity spectrum is discrete, at least
when considered with respect to the Hamiltonian generating global
time translations.

In \cite{dWFGK} a coordinate transformation was given for the Poincare patch
of
AdS$_5$ from written as a warped product with an AdS$_4$ Poincare patch to
the standard form. One 
can extend this coordinate transformation to the full global solution:
\eqn{GlobalAdSfour}
{ds^2 = L^2 \cosh^2 (\frac{c-r}{L}) \; ds_4^2 \; - \; dr^2 }
where
\eqn{GlobalAdSf2}
{ds_4^2 = \frac{1}{\cos^2 (\rho)} \; \left ( dt^2 - d\rho^2 - \sin^2 (\rho) 
\; d \Omega_2^2 \right). }
Applying the following change of variables
\eqn{VariableChangeglobal}{\seqalign{\span\TR & \quad \span\TR}{
     \tan (\tilde{\rho}) = \coth (\frac{c-r}{L})  \; \tan (\rho)
     \ , \quad
    \cosh (\tilde{r}) = \frac{\cosh (\frac{c-r}{L})}{\cos (\rho)}
      \ , \cr
     \tilde{t} = t \ , \quad
      \tilde{\Omega}_2 = \Omega_2 \ 
  }}
we end up with
\eqn{GlobalAdSfive}{
ds^2= L^2 \left ( 
\cosh^2 (\tilde{r}) \; dt^2 - d\tilde{r}^2 - \sinh^2 (\tilde{r})
d \tilde{\Omega}_3^2 \right ) }
where $\tilde{\Omega}_3$ is the 3-sphere metric on $\tilde{\rho}$ and
$\tilde{\Omega}_2$, $d \tilde{\Omega}_3^2=d \tilde{\rho}^2 + \sin^2 (
\tilde{\rho})
\tilde{\Omega}_2^2$. 
This can be recognized as global AdS$_5$, see for
example \cite{magoo}\footnote{Note that while $\rho$ runs from 0 to 
$\frac{\pi}{2}$ and $r$ from $- \infty$ to $\infty$, 
$\tilde{\rho}$ runs from 0 to $\pi$ and $\tilde{r}$ from
0 to $+\infty$. When we take $r$ to zero, $\tilde{\rho}$ has to go
to $\frac{\pi}{2}$. As $r$ goes negative, on the other side $\tilde{\rho}$
now explores the region between $\frac{\pi}{2}$ and $\pi$}. Using
this change of variables, it is easy to see, that an $r=const.$ surface
is embedded in global AdS$_5$ as indicated in Figure~\ref{slice}.

Note that this implies an interesting realization of holography.
Gravity in the bulk of our space should be dual to a theory
living on the full boundary, which is composed of two pieces,
the AdS$_4$ brane and the $S^3/Z_2$ half-boundary of the original
AdS$_5$,
with two different field theories living on them. They touch along their
common boundary. According to AdS/CFT lore, any theory in AdS can
be completely encoded in its boundary data, so that in the end all 
AdS$_4$ bulk
physics should be captured by the CFT on the disk with 
the brane physics encoded in the boundary data of the disk. 
Instead of a CFT with momentum somehow cutoff,
we get a CFT with spatial cutoff
(that is formulated on a disk). The reduction of the symmetry
from $SO(4,2)$ to $SO(3,2)$ corresponds to the well known fact that
this is the subgroup of conformal transformations that leaves a boundary
invariant.

\subsection{The Discrete Spectrum in the Volcano Potential}
We now solve the quantum mechanics problem with the box-like
volcano potential \volcano\ . In order to solve the spectrum with  a delta
function at the bottom of a box, we first find the eigenfunctions
in the box, and then look for the right linear combinations to satisfy
the jump imposed on the first derivative of the wavefunction due to the
delta function.

We first consider the zero tension case, where formally
the delta contribution vanishes\footnote{If one actually wants
to analyse the spectrum of an almost tensionless brane, the
spectrum will be the same, but every other mode is
projected out by the boundary condition at the brane.
By just omitting the delta function we are really looking
at `no brane' instead of a zero tension brane.}.  The turn around
point $c$ in the warp factor \deSitterWall\ goes to zero, the solution
becomes just $A=\log \left( L \sqrt{\Lambda} \cosh{\frac{r}{L}} \right )$,
and we recover pure AdS$_5$. The parameter $z_0$ in \volcano\
becomes $\sqrt{|\Lambda|}
 z_0=\frac{\pi}{2} $ and the resulting potential is just
\eqn{volcanoinads}{\eqalign{
    \quad V(z) = \frac{-9 (-\Lambda)}{4} + \frac{15}{4} \frac{
(-\Lambda)}{
\cos^2 \left ( \sqrt{-\Lambda} z \right )} 
  }}
with $z$ running from $-\frac{\pi}{2}$ to $\frac{\pi}{2}$.
Rescaling $z \rightarrow \sqrt{-\Lambda }z$,
Schr\"odinger's equation with energy eigenvalue $\frac{E}{-\Lambda}$ 
in the potential \volcanoinads\ can be solved
explicitly and the resulting wavefunction is
\eqn{wavefunction}{\seqalign{\span\TR  & \quad \span\TR}{
\psi(z) =& c_1 \cdot \; _2F_1(-\frac{3}{4} -
\frac{\sqrt{9 + 4E}}{4}, 
\frac{5}{4} -\frac{\sqrt{9 + 4E}}{4};
1-\frac{\sqrt{9 + 4E}}{2}; \frac{1}{\cos^2 z}) \; \cos^{\frac{\sqrt{9 + 4E}}{2}}
z \; + \cr &c_2 \cdot \; _2F_1 (-\frac{3}{4} +
\frac{\sqrt{9 + 4E}}{4},
\frac{5}{4} +\frac{\sqrt{9 + 4E}}{4};
1+\frac{\sqrt{9 + 4E}}{2}; \frac{1}{\cos^2 z}) \; \cos^{\frac{-\sqrt{9 + 4E}}{2}}
z.}}
In order to not diverge on either side of the box we need
$c_2=0$ and
\eqn{eigenvalues}{ E=n \; (n+3) \; \mbox{   where   } \; n=1,2,3,\ldots . }

This spectrum indeed agrees with the spectrum of gravity fluctuations
of pure AdS$_5$, as can be seen to be the case by group-theoretic
considerations. Unitary representations of the AdS$_5$ group are classified
(for a recent discussion see e.g. \cite{frondsdalzaff}) by 
the eigenvalue of the global time
Hamiltonian $E_0$, which is the generator of the $SO(2)$ subgroup
of $SO(4) \times SO(2) \subset SO(2,4)$,
and $SO(4)$ spins $j_1$ and $j_2$ for the groundstate
from which the infinite dimensional representation is built
by raising operators. The massless graviton (where masslessness
is defined via the smaller number of polarization states) has
$E_0=4$ and $j_1=j_2=1$.
The analysis of the analog quantum mechanics amounts to decomposing
the $SO(4,2)$ representation under the $SO(3,2)$ subgroup that is
manifest in the warped product based on AdS$_4$. As in 5d, these are
classified by $E_0$ and $s$, where $E_0$ in 4d is identified with
$E$ in 5d (the $SO(2)$ subgroup corresponding
to global time translations of $SO(4,2)$ is the same as
in $SO(3,2)$) and $s$ is the $SO(3)$ spin obtained by tensoring
together $j_1$ and $j_2$. The Casimir now reads
\eqn{Casimir4d}{C_2(E_0,s) = E_0 (E_0-3) + s (s +1)}
and a massless representation obeys
\eqn{massless4d}{E_0 = s+1.}
So the apropriate definition of mass for a spin 2 representation is
\eqn{massinads4}{m^2=C_2(E_0,2) - C_2(3,2) = E_0 (E_0 -3)}
The single massless graviton decomposes into a tower of spin 0,1 and 2
representations in four dimensions. The masses of the spin 2 excitations
are
\eqn{ads5spectrum}{ m^2 = E_{0,n} (E_{0,n}-3) \; \mbox{ with } \;
E_{0,n} = 4,5,6,\ldots }
which obviously agrees with the spectrum
\eigenvalues\ found via the Schr\"odinger
equation.

Upon including the brane the solution in terms of the
linear combination of hypergeometric functions
 becomes quite complicated and we chose to study it 
numerically\footnote{
For an analytic treatment see \cite{andre}.}. 
A more detailed analysis of the spectrum
will appear elsewhere. Let us briefly exhibit the crucial features of
the spectrum as obtained via a numerical solution of
Schr\"odinger's equation using the shooting method.

In order to find the eigenvalues and wavefunctions for
non-zero tension, we use 
\eqn{startshoot}{\frac{\psi'}{\psi} = -3 \tan(z_0)}
as required by the delta function in the potential 
as the start condition for wavefunction on the brane and then numerically
solve Schr\"odinger's equation outwards. For generic energy values, the
would-be wavefunction will diverge at the boundary.
Eigenvalues are the values of the energy for which
this expression goes to zero at the boundary.
Of course, this is never really zero in any numerical approximation.
The shooting method relies on the fact that (for any reasonable potential)
the would be wavefunction diverges to (say) +infinity if you are below an
eigenvalue and then will diverge to -infinity above the eigenvalue (and
then vice versa for the next eigenvalue).  
If one plots the
wavefunction at the boundary as a function of energy, the
zeroes of this function are the eigenvalues.
\begin{figure}
   \centerline{\psfig{figure=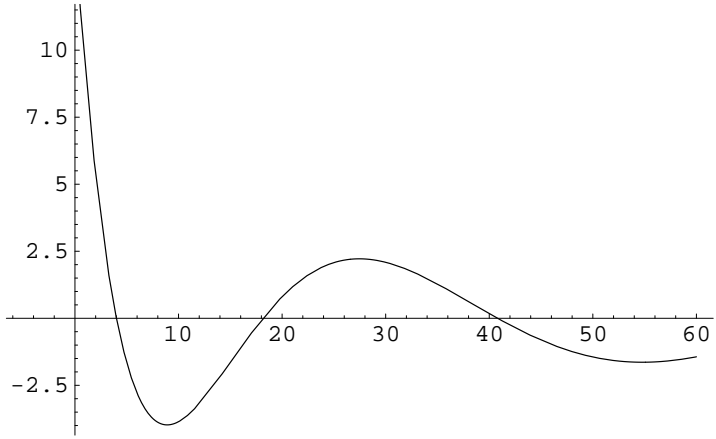,width=3.7in}}
    \caption{Spectrum at zero tension, $\sqrt{|\Lambda|} z_0 = \frac{\pi}{2}$.}
\label{fevzero}
  \end{figure}
\begin{figure}
   \centerline{\psfig{figure=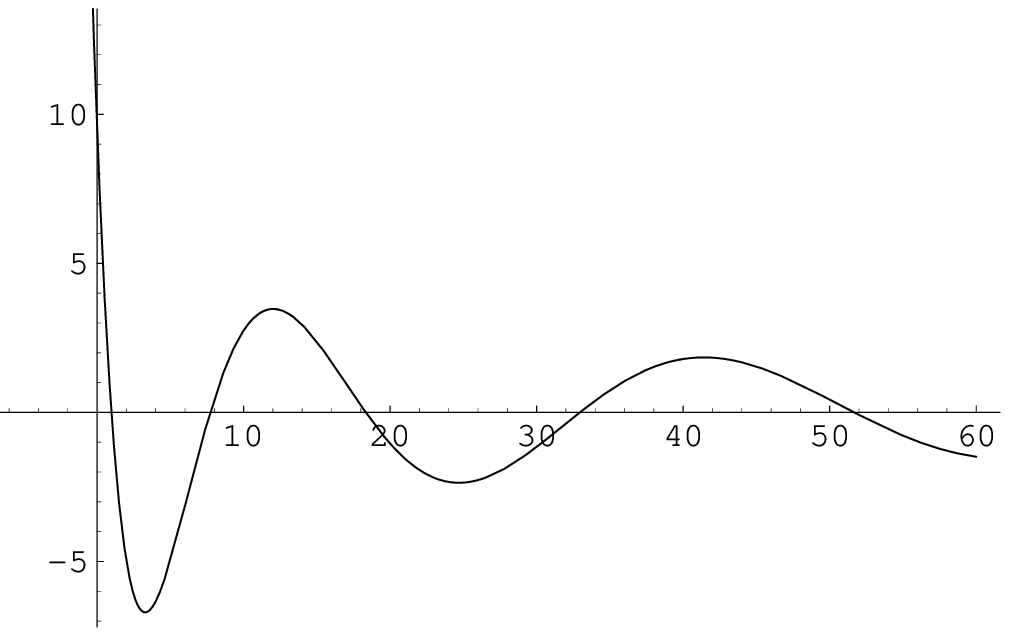,width=3.7in}}
    \caption{Spectrum at $\sqrt{|\Lambda|} z_0=\frac{\pi}{2} -0.7$.}
\label{fev7}
  \end{figure}
\begin{figure}
   \centerline{\psfig{figure=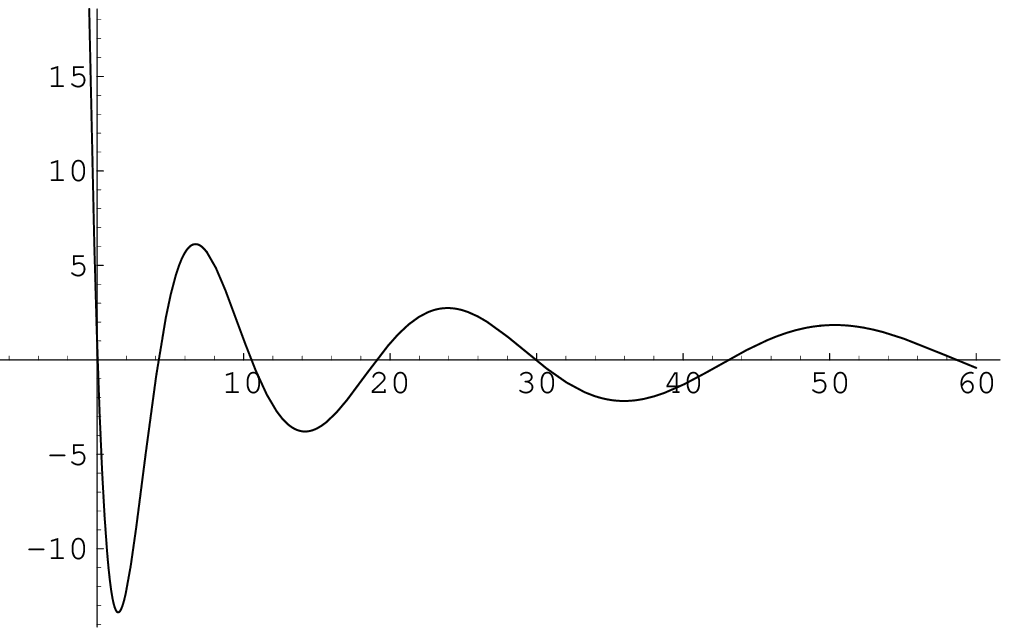,width=3.7in}}
    \caption{Spectrum at $\sqrt{|\Lambda|} z_0=\frac{\pi}{2}-1.4$.}
\label{fev14}
  \end{figure}

Figures~\ref{fevzero},~\ref{fev7},~\ref{fev14} 
show the spectrum of low lying modes
as we increase the tension. For small tension, we keep every other
mode of the pure AdS$_5$ analysis. As we increase the tension, one mode
comes down to mass\footnote{
A more detailed numerical analysis \cite{matthew} or
an analytic treatment relying on the factorization of the hamiltonian
\cite{andre} both show that the actual behavior for the
almost zero mode is $m^2 \sim \frac{3}{2} \Lambda^2$.} 
 $m^2 << \Lambda$, while all the other modes
stay at $m^2 = {\cal O} (\Lambda)$. As we approach the critical tension,
the potential becomes ``two copies'' of the pure AdS$_5$ potential pasted
together along a delta function. Indeed we have all the heavy
modes of the AdS$_5$ analysis at $m^2=(n-3)n\Lambda$ and in addition
a very light  mode trapped in the delta function. In the critical
limit the trapped very light mode becomes the trapped massless graviton,
while the densely spaced excited modes become the continuum of KK
modes.
\begin{figure}
   \centerline{\psfig{figure=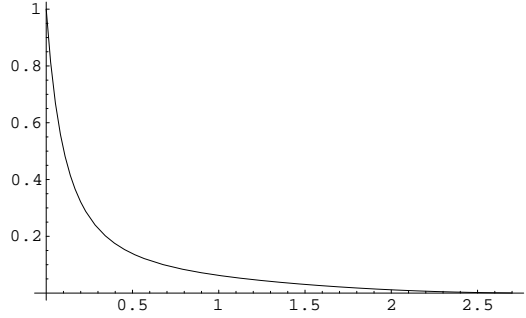,width=2.7in}}
    \caption{Wavefunction of the almost massless mode, E=0.0419 at 
$\sqrt{|\Lambda|} z_0=\frac{\pi}{2} -1.4$.}
\label{almostzero}
  \end{figure}
\begin{figure}
   \centerline{\psfig{figure=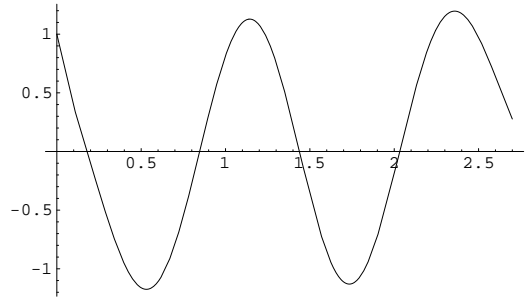,width=2.7in}}
    \caption{Excited mode at $\sqrt{|\Lambda|} 
z_0= \frac{\pi}{2} -1.4$ and E=30.}
\label{thirty}
  \end{figure}
Figures~\ref{almostzero},~\ref{thirty} show the wavefunctions of the
almost massless mode and of a highly excited mode. Comparision
with the shape of the zero mode and the KK-modes in the RS
scenario show indeed that in the flat
space limit, the former approaches the zero mode
localized on the brane,
while the latter  approaches a  highly excited KK mode, oscillating 
without feeling the brane at all. The low KK modes also look as expected:
oscillating in the bulk but suppressed on the brane.

As we decrease the tension, the wavefunction of the 
almost zero mode gets broader
and loses its amplitude enhancement over the KK modes,
as is studied in detail in \cite{matthew}.
Therefore, we would never actually be able to
see four-dimensional  AdS gravity and at the same time a significant
deviation due to the
graviton mass. For small $\Lambda$, the mass is tiny,
and by the time we reach distance scales of order $1/m$ where
the mass of the graviton should become significant, all
excitations have died off anyway due to the gravitational potential
of AdS. When we go to $\Lambda \rightarrow \frac{1}{L^2}$,
which is the maximal value we can reach for very small tension branes,
the mass of the almost zero mode goes like $4 \Lambda$ and hence
one should now see it contributing on the same footing as the
4d curvature effects. However since in this limit, the wavefunction
is no longer bound to the brane and the almost zero mode has 
a mass comparable to the other excited modes,
it really becomes one of the KK-modes and one no longer has
four-dimensional gravity at all.

\section{The non-TT Modes}

So far we have only analyzed the spectrum of TT fluctuations.
We have seen that in the AdS case, we get a massive spin 2 almost
zero mode and the usual massive KK tower. This
naively seems inconsistent, since the massive mode has
extra polarization states that should contribute to the propagator.
However, we expect a smooth flat space limit.
In order to complete the analysis,
we have to study the non-
transverse-traceless modes as well. We will argue
that although the additional modes can be gauged away up to a brane
bending effect, they nonetheless contribute
to the propagator,   effectively reducing
the degrees of freedom from 5 to 2. This will
be discussed further in Ref. \cite{withami}.

We first analyze the 
general form of a solution to the equations
of motion and then compare this with
the residual gauge transformations.
We still write the perturbed metric as
\eqn{PerturbedMetricrep}
{ds^2 = e^{2 A(r)} (g_{ij} + h_{ij}) \, dx^i dx^j - dr^2
\,,}
in axial gauge
\eqn{AxialGaugerep}{ h_{\mu 5} = 0 \,,} where $\mu = 0,1,2,3,5$,
but do not restrict to TT components only.
We first exhibit the general form of
the solution to the bulk equations of motion.

The first order equations of motion require\footnote{
In the $ij$ equations we used the zeroth order equations of motion
to eliminate the terms involving the stress energy tensor
on the right hand side of Einstein's equations as in
\cite{dWFGK,dWF}.} vanishing of (where $h=g^{ij}h_{ij}$):
\eqn{fullfluctuations}{\eqalign{
55: & \quad (e^{2A} h')' \cr
5i: & \quad -\frac{1}{2} \d_i h' + \frac{1}{2} \d^k h_{ik}' \cr
ij: & \quad e^{2A} (\frac{1}{2} h_{ij}'' + 2 A' h_{ij}') -
\frac{1}{2} \d^2 h_{ij} + \frac{1}{2} g_{ij} e^{2A} A' h' - \cr
    & \quad \frac{1}{2} \d_i \d_j h +
\frac{1}{2} ( \d^k \d_i h_{jk} + \d^k \d_j h_{ik})+
3 \Lambda h_{ij}
\ .
}}
In order to manipulate these expressions, the following identities
for covariant derivatives in dS and AdS are useful:
\eqn{identities}{\eqalign{
R_{ijkl} =& \Lambda (g_{il} g_{jk} - g_{ik} g_{jl}) \cr
R_{ij}  = g^{jl} R_{ijkl} =& - 3 \Lambda g_{ij} \cr
\d_i \d_j \d_k f -
\d_j \d_i \d_k f =& \Lambda (g_{jk} \d_i - g_{ik} \d_j) f \cr
\d_i \d_k \d^k f - \d^k \d_i \d_k f =& 3 \Lambda \d_i f \cr
\d_i \d_j \d^k \d_k f - \d^k \d_k \d_i \d_j f =&   (8 \Lambda \d_i \d_j
- 2 \Lambda g_{ij} \d^2 ) f \cr
\d^k \d_i \d_j \d_k f - \d^k \d_k \d_i \d_j f =& (- \Lambda g_{ij} \d^2 f
+ \Lambda \d_i \d_j f) \ .
}} From
the 55 equations it follows immidiately that $h_{ij}$ can be written
as
\eqn{afterff}{h_{ij} = h_{ij}^{TT} + G \; f_{ij} + d_{ij} }
where $h_{ij}^{TT}$ are the TT components of the
fluctuation which we studied in great detail before, $f_{ij}$
and $d_{ij}$ are functions of the 4d space-time only
and 
$G$ is defined as\footnote{This of course only defines $G$ up to a constant,
which can be absorbed into $d_{ij}$. To be precise, we will always
choose 0 as the lower bound on the integral.}
\eqn{G}{ G = \int^r e^{-2A(\tilde{r})} d\tilde{r}. }
Now from the $i5$ we find
\eqn{fromif}{ \d_i f^k_k = \d^k f_{ik}. }
Using as an ansatz
\eqn{ansatzforf}{f_{ij} = \d_i \d_j \alpha - g_{ij} \beta}
we see that \fromif\ implies
\eqn{afterif} { f_{ij} = (\d_i \d_j - \Lambda g_{ij}) \Phi. }
$\Phi$ is what we would like to identify as the radion field.
Plugging this into the $ij$ equation, one finds after a tedious calculation
using \identities\ and the following property of
our warpfactors
\eqn{warpident}{
A' + G \Lambda = - \frac{\lambda}{3} }
that the full system is solved by
\eqn{hsolved}{ h_{ij} = h^{TT}_{ij} + 2 G \; (\d_i \d_j - \Lambda g_{ij})
\Phi - \frac{2}{3} \lambda g_{ij} \Phi }
where $\Phi$ has to satisfy the following equation of motion:
\eqn{radioneom}{ \d^2 \Phi = 4 \Lambda \Phi }
and the TT modes $h^{TT}_{ij}$ are determined by the analog
quantum mechanics we analyzed before. The boundary condition at
the brane demands
\eqn{boundary}{h_{ij}'=0= h^{TT'}_{ij} + 2 e^{-2A} (\d_i \d_j - \Lambda
g_{ij} ) \Phi .}

Now let us proceed to analyze the residual gauge transformations
that remain after fixing axial gauge.
Under gauge transformations that take $x^{\mu} \rightarrow x^{\mu}
+ \xi^{\mu}$, the fluctuations transform as
\eqn{gaugetrafosfull}{\eqalign{
h_{55} \rightarrow & \quad h_{55} - 2 e^{-2A} \xi^{5'} \cr
h_{ij} \rightarrow & \quad h_{ij} + (\d_i \xi_j + \d_j \xi_i) +
2 A' g_{ij} \xi^5 \cr
h_{i5} \rightarrow & \quad h_{i5} - e^{-2A} \d_i \xi^5 + g_{ij} \xi^{j'}
\  }}
where we pulled out a factor of $e^{2A}$ from the $h_{55}$ and $h_{i5}$
component as we did for the $h_{ij}$ fluctuation in \PerturbedMetricrep\
and $G$ is defined as in \G\ above.
In order to preserve the axial gauge \AxialGaugerep\ we require that
$h_{55}$ and $h_{i5}$ stay zero, leaving us with the residual gauge
transformations
given by
\eqn{residual}{\eqalign{
\xi_5 =& \quad \epsilon^5(x) \cr
\xi_i =& \quad G \d_i \epsilon^5 + \epsilon_i(x) \ . }}

As in \cite{katz},
the $\epsilon^i$ gauge transformations are needed in order
to gauge away the longitudinal components of the graviton
fluctuations, which we already neglected by our ansatz for
$h_{ij}$.
Using \warpident\ one can see that the residual gauge transformations
\residual\
preserving axial gauge
can indeed be used to gauge away the radion component $\Phi$ in the
full solution \hsolved\ by choosing $\epsilon^5 = -\Phi$.

In the presence of sources on the brane, the $\epsilon_5$ gauge
transformation leads, as in \cite{katz}, to a bent brane in
the new gauge and to a brane bending
contribution to the propagator, proportional to
the propagator of a scalar field of mass $4\Lambda$.
This scalar
is clearly not one of the physical modes, since we can gauge it away everywhere
where there is no stress energy. Hence it does not correspond
to a propagating ghost. In this respect our setup differs crucially
from the quasilocalized gravity model of \cite{GRS} that was
analyzed in a very similar fashion 
in \cite{oldkogan,erlich,terningRG,nyu,nyu2}.
There, due to the additional boundary condition on the second
brane the radion cannot be gauged away \cite{zaffGRS} and
represents a real ghost.
In the critical tension limit, the bending contribution
is precisely what is needed
to smoothly match on to the standard massless 4d graviton propagator.
The absence of a Veltman Van Dam discontinuity \cite{vvd,zahkarov},
as
the mass of the graviton goes to zero is consistent with the findings
of \cite{mp}
\footnote{After completing this work, we became aware of \cite{oxford},
which obtains a similar result to \cite{mp}.}
that states that no such discontinuity exists in AdS space 
as long as the mass $m^2$ goes to zero faster than $\Lambda$.
 This shouldn't come as
a surprise, since 
we preserved the gauge invariance corresponding to 4d diffeomorphisms
despite the fact that we are dealing with a massive graviton.
As will be explained in \cite{withami}, these additional gauge invariances
can still be used to eliminate 3 of the polarization modes.

\section{Conclusions}

In this paper, we have demonstrated that localization
of gravity applies, even for the $AdS_4$ brane
for which there is no normalizable zero mode. Although
expected from the vantage point of locality
and holography, it is interesting  that a massive
graviton generates four-dimensional gravity.

Our result differs substantially
from quasi-localization models, that also had
a turn-around in the warp factor that modifies
gravity on long distance scales. In the
existing theories, there was a ghost
when four-dimensional gravity is reproduced.
 Furthermore, those theories
required unphysical matter in order
to violate the C-theorem.

Our theory should have interesting implications for
holography. 
A complete holographic description of the setup 
can be given in terms of a CFT living on the disk that
is the left over part of the true AdS$_5$ boundary.
In this description, all of the physics on the brane is reduced
to correlators on the common boundary of the disk and AdS$_4$.
This description is however not appropriate to study local
physics on the brane.
For latter purpose it should be more useful to split the
bulk excitations into two sets, one dual to a CFT on
the true boundary, the other one to a CFT on the brane,
but the precise procedure is unclear.
The mass of the almost zero mode
is attributable to the long distance behavior
of the warp factor which is sensitive to boundary
physics; presumably the mass can be understood
as a consequence of the boundary CFT. Understanding
how the theory on the brane is reproduced by CFT physics 
remains a challenge.

The much more general phenomenon of localization
demonstrated in this paper might have other
implications.  Compactification of extra
dimensional spaces is not essential to reproduce
four-dimensional physics; a region of space which localizes
gravity is sufficient. Furthermore,  the current no-go theorems
about localization in supergravity theories e.g. \cite{
maldacena,cvetic,kallosh}
relying on the asymptotic behavior of the geometry
do not necessarily apply. It is important to
know whether this physics can be realized in string theory.

\section*{Acknowledgements}
We want to thank M. Aganagic,
O. Aharony, 
N. Arkani-Hamed,
A. Chamblin, O. DeWolfe, 
D. Freedman,   M. Gremm, 
  A. Katz, A. Miemiec, A. Naqvi, 
L. Rastelli, M. Schwartz, and B. Wald for 
useful discussions. We especially thank M. Porrati, both
for discussions and for sharing his results \cite{mp}.
We are also grateful to the organizers of the conferences at which
much of this work has been previously reported \cite{talks}.
Our work was supported by the U.S. Department of Energy under
contracts \# DE-FC02-94ER40818 and \# DE-FG02-91ER4071.

\bibliography{lisa}
\bibliographystyle{ssg}

\end{document}